\documentclass[journal]{IEEEtran}
\usepackage{graphicx}
\usepackage{subfigure}

\begin{document}

\title{Modeling Probability of Path Loss for DSDV, OLSR and DYMO above 802.11 and 802.11p}

\author{S. N. Mohammad$^{1}$, S. Wasiq$^{1}$, W. Arshad$^{1}$, N. Javaid$^{1}$, S. Khattak$^{2}$, M. J. Ashraf$^{3}$\\\vspace{0.4cm}
        $^{1}$COMSATS Institute of Information Technology, Islamabad, Pakistan.\\
        $^{2}$Abasyn University, Peshawar, Pakistan.\\
        $^{3}$Islamabad Electric Supply Corporation, Islamabad, Pakistan.
     }

\maketitle

\begin{abstract}
 This paper presents path loss model along with framework for probability distribution function for VANETs. Furthermore, we simulate three routing protocols Destination Sequenced Distance Vector (DSDV), Optimized Link State Routing (OLSR) and  Dynamic MANET On-demand (DYMO) in NS-2 to evaluate and compare their performance using two Mac-layer Protocols 802.11 and 802.11p.
 A novel approach of this work is modifications in existing parameters to achieve high efficiency.
 After extensive simulations, we observe that DSDV out performs with 802.11p while DYMO gives best performance with 802.11.
\end{abstract}


\IEEEpeerreviewmaketitle

\section{Introduction}

\IEEEPARstart{M}{obile}  Ad-hoc Network (MANETs) is collection of independent mobile users taken as mobile nodes that communicate through wireless links.
 The creation of network protocols for these network topologies is a complex issue.
  Vehicular Ad-hoc Networks (VANETs) are distributed, Self-assembling communication networks that are made up of multiple autonomous moving vehicles, and peculiarized by very high node mobility. The major purpose of VANETs is providing protection and ease to the travelers.

There is no single unique protocol that is convenient for all networks impeccably. The protocols have to commensurate to network's unique characteristics, such as density, scalability and the mobility of the nodes. The routing protocols subdivided into table driven and on-demand based on the behaviour of protocols . In table driven, proactive protocols are based on periodic exchange of control messages and maintaining routing tables. Each node maintains complete information about the network topology locally. However, the reactive protocol tries to discover a route only on-demand, when it is necessary. It usually takes more time to find a route compared to a proactive protocol. Our stimulation work is based upon comparison of three protocols in MANETs and VANETs named as OLSR [1], DSDV [2] and DYMO [3].


\section{Related Work and Motivation}
Several papers have been published regarding the comparison of routing protocols in different simulation scenarios. The comprehensive modeling for VANETs is been done in Khabazian et.al [4-5]. This work is been improved and more generalized in section III.
The comparison for AODV and DSR is carried out in realistic urban scenario with varying Node mobility and Vehicle density to observe the behavior of both protocols [6]. In this study, modified version of OLSR has been discussed. In accordance to this work we made some modifications in all other routing protocols which is shown below. The paper also shows comparison of DSR and DSDV with four different mobility models i.e., Random Waypoint, Group Mobility, Freeway and Manhattan model is presented in [7].
A few studies are carried out to evaluate the performance of different routing protocols in VANETs for some scenarios [8].

The works in [9-13], study the most widely experimented and frequently used protocols for our study; three from reactive or on-demand class: AODV, DSR, DYMO, and three from proactive or table-driven class DSDV, FSR, OLSR.

 .



\section{Mobility Model for VANETs}

\IEEEPARstart{I}n this section, we will present the probability distribution function $pdf$ of the distance of a node from the strip segment. In [5], movement of each node is taken as a function of time consisting of a sequence of random intervals called mobility epochs. The epoch durations of each node is exponentially distributed with mean $\frac{1}{\beta}$. During each epoch a node moves at a constant speed chosen independently from a normal distribution with mean $\mu$ and variance $\sigma$. The distance of a node as a function of time from the highway segment, X(t), follows a normal distribution with the $pdf$, in eq. (1).

\begin{eqnarray}
{b_{x(t)}}(r)=\frac{1}{\sqrt{2{\pi}\Theta_{x(t)}}}\,e^\frac{-(r-\varepsilon_{x(t)})^2}{2\Theta_{x(t)}}
\end{eqnarray}
where mean $\epsilon_{x(t)}={\mu t}$, \\
and variance, $\Theta_{x(t)}=a(\beta t-1+e^{-\beta t})$,with $a=\frac{2\sigma^2}{\beta^2}$. \\
 Let $B_{x(t)}(r)$ denote the corresponding cumulative distribution function (cdf),therefore,

\begin{eqnarray}
B_{x(t)}(r)=Pr(X(t)\leq r)=\int_{0}^{r}{b_{x(t)}y dy. }
\end{eqnarray}
From eq. 2, we may note that the mean and variance increase as the time increases. As a result, pdf shifts towards the right, its maximum value shrinks and gradually becomes more flat. Thus pdf approaches to a uniform distribution with increasing time.

\begin{figure*}[t]
 \centering
{\includegraphics[height=11 cm,width=14 cm]{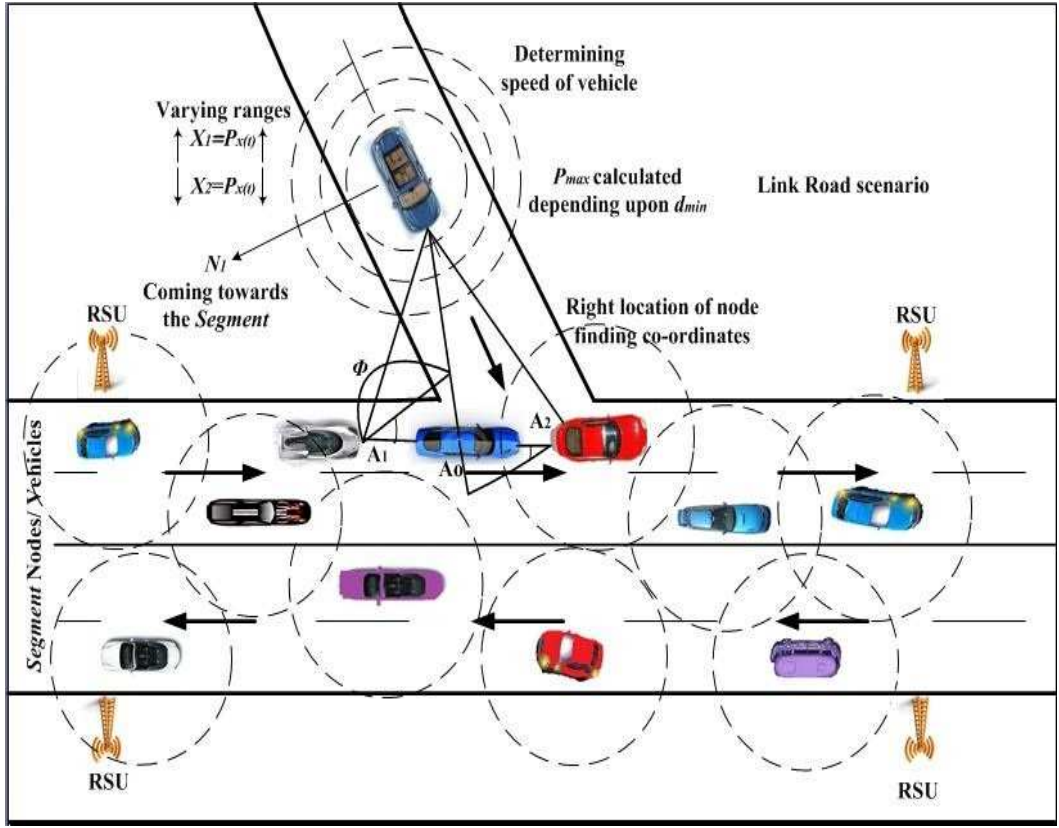}}
\caption{System Model}
\end{figure*}


Let $X_n(t)$ denote as random mobility distance of the node given that there has been $n$ epochs during the time interval t, and $X_{n(t)}=\sum_{i=1}^{n}x_i=\sum_{i=1}^{n}s_i t $ where $s_i$ is the speed of a node during $i'th$ epoch. The mean and variance of $X_n(t)$ are given by,\\
\begin{eqnarray}
m_{x_{n}(t)}=nE(x_i)=nE(s_i t_i \mid n epochs)=\mu t
\end{eqnarray}
Therefore, $m_{x_{n}{(t)}} = \frac {t}{\beta}$.
If a random variable $X_n(t)$ has the expected mean value $\mu = E[xi]$, then variance of random variable $X_n(t)$ is typically designated as $var(x)$,

\begin{eqnarray}
var_{x_{n}(t)} = E[(xi -\mu)^2]
\end{eqnarray}

The variance is the expected value of the squared difference between the variable's realization and the variable's mean.
After expanding eq.(4), we get eq.(5) which shows mnemonic for the expression, "mean of square minus square of mean".
Therefore, we can write as\\

\begin{eqnarray}
var_{x_{n}(t)}=n[E({x_i})^2-E^2{(x_i)}]
\end{eqnarray}

As in [4] the first two moments for the few initial values of the number of epochs are of $E(x_i)=\frac{\mu t}{n}$ and $E(x^{2}_i)=E(s^{2}_i)E(t^2_i\mid n epochs)$ the variance as mentioned in [4] is then calculated as,

\begin{eqnarray}
var_{x_{n}(t)}=at-(bt+\frac{a}{\beta})(1-e^{-\beta t})
\end{eqnarray}

In [5] numerical study epoch rate and the standard deviation of the nodes speed are set to constant values of $\beta=1$ and $\sigma=3$ respectively. Now by using the above mention values, variance is formulated as\\
\begin{eqnarray}
var_{x_{n}(t)}=18t-(t+18)(1-e^{-t})\\
var_{x_{n}(t)}=18(t-1+e^{-t})
\end{eqnarray}

Let the probability of communication for any node the strip of the segment be $p_{x(t)}$.therefore,\\

\begin{eqnarray}
p_{x(t)}=\frac{\gamma}{\omega}
\end{eqnarray}

where $\omega={\sqrt{2{\pi}{var_{x_{n}(t)}}}}$ and $\gamma=e^\frac{-(r-t)^2}{2 var_{x_{n}(t)}}$ $r\varepsilon(0,200)$\\
\begin{eqnarray}
P_{x(t)}(r)=\int_{0}^{r}{p_{t}z dz. }
\end{eqnarray}
Taking the distance of the nodes as in [5] and simulation time written in Table[1]. We deduce that the probability $P_{xn(t)}$ of
node $N1$ as shown in Fig. 1 increases as distance is greater with the strip of segments.

\begin{figure*}[t]
 \centering
\subfigure{\includegraphics[height=5cm,width=8 cm]{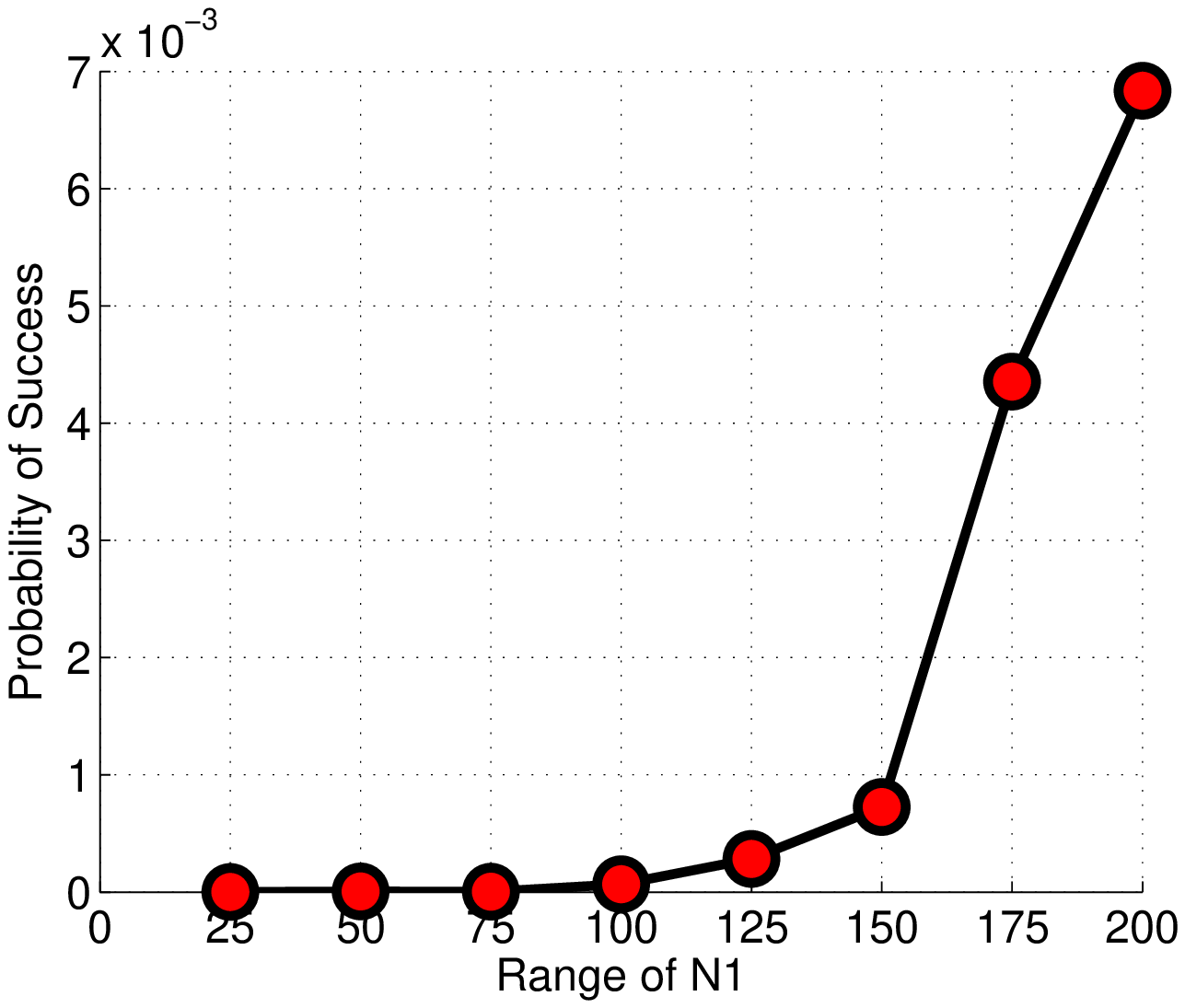}}
\subfigure{\includegraphics[height=5 cm,width=8 cm]{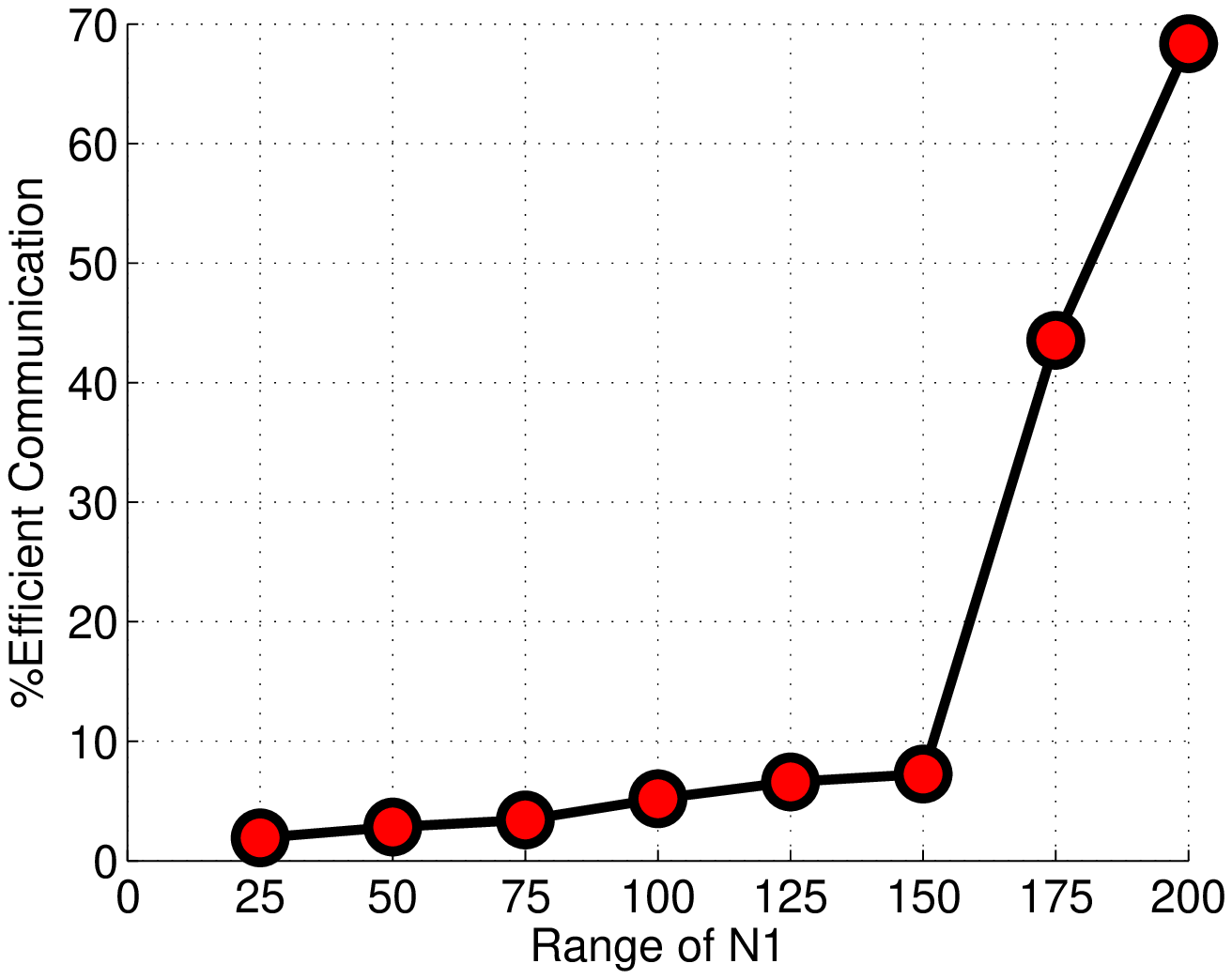}}
\caption{Probability and Efficiency of successful communication}
\end{figure*}

Now calculating the efficiency of these probability, showing how much efficient is the communication between node and
strip of the segment. As
\begin{eqnarray}
\eta = p_{xn(t)} * 100
\end{eqnarray}


Now, $N1$ as shown in the fig.1 is coming towards the
strip of the segment establishing communication. We consider
$N1$ as a sender node and strip of the segment
as the receiving nodes such that $N1$ sends some messages
received by $Segment$ nodes.
 Denoting
the distance as $d(N1,Segment)$ between $N1$ and $Segment$.
Let denote as $P_{N1}$ the sending power of the node N1. As
in VANETs, vehicles act as nodes so these nodes $N1$ and Segment
nodes are also vehicles. Each vehicle equipped with VANET
device and can receive and relay other messages through the
wireless network. These devices contain the network sensors
involving antennas in them. With an isotropic antenna of gain
$G$ and for $d(N1, Segment)$ sufficiently large, the $Segment$
nodes will receive a power $P_{Segment}$ which is expressed as

\begin{eqnarray}
\frac{P_{N1}}{P_{Segment}}=\frac{(4\pi)^2 d^{2}(N1, Segment)}{\lambda^{2} G_{N1} G_{Segment}}\\
P_{Segment} = P_{N1} \frac{\lambda^2 G_{N1} G_{Segment}}{16{\pi}^2 d^{2}(N1, Segment)}
\end{eqnarray}

$\lambda$ is the wavelength of the transmitted signal. Transforming
eq. (12) in terms of $dB$ by using logarithmic scale.

\begin{eqnarray}
10\log_{10}\frac{P_{N1}}{P_{Segment}}=10\log_{10} \frac{(4\pi)^{2} d^{2}(N1, Segment)}{\lambda^{2} G_{N1} G_{Segment}}
\end{eqnarray}

By taking the gain of the link from $N1$ to $Segment$ as

\begin{eqnarray}
G_{d(N1,Segment)} = G_{N1} G_{Segment}\frac{\lambda^2}{16\pi^2}
\end{eqnarray}

the maximal power $P_{max} d(N1, Segment)$ from
the sender can also be re-written as:

\begin{eqnarray}
P_{max}d(N1, Segment)=\frac {P_{N1}G_{d(N1,Segment)}}{d^{2}(N1, Segment)}
\end{eqnarray}

By taking into account signal attenuation send by $N1$, the
power received by $Segment$ node is smaller than $P_{max}$

\begin{eqnarray}
P_{Segment} = \alpha P_{max}(d(N1, Segment))\,\,\,\,\,\, 0 \leq \alpha \leq 1
\end{eqnarray}

Where $\alpha$ depends on several parameters like congestion of traffic,
collision and interference between vehicles communication.
We denote $d_{min}$ the minimal distance between $N1$ and
$Segment$ nodes. We can define the maximal received power
$P_{max}$

\begin{eqnarray}
P_{max} =\frac {P_{N1}G_d(N1,Segment)}{d^2_ {min}}
\end{eqnarray}

\section{Path Loss Model for VANETs}

The path loss model we use to evaluate the distance between
$N1$ and $Segment$ is the Friis Free Space Path Loss Model,
which represents the signal attenuation when
line of sight between the $N1$ and $Segment$ nodes is clear. This model
stipulates that:
\small
\begin{eqnarray}
PL(d)[dB] = 2PLfs(do)[dB] + 10nlog_{10}[\frac{d(N1, Segment)}{do}]
\end{eqnarray}
\normalsize
PL is the path loss, PLfs is the path loss within a free
space environment. $d_o$ is a define as received power with
respect to reference point while $n$ is the path loss exponent which
represents the increase of path loss with the increase of the
distance between the $N1$ and $Segment$ nodes.

To estimate distances between $N1$ and $Segment$ nodes at
different positions of the segment, we use the Friis model.
From their distances an angle can be deduce between the two
nodes, as illustrated in Fig. 1, where the $Segment$ nodes
localize the $N1$ sender by determining the angle $\Phi$ between
them. For that, the receiver start evaluating the received
strength from the sender, at positions $A1$ and $A2$. The distance
between these positions is $L$. Then, using the Friis path loss
model, the receiver can evaluate the distances x1 and x2 at
positions $A1$ and $A2$.

We suppose that the distance between the $N1$ and $Segment$
is equal to the average between the distances $x1$ and $x2$, where
$x_1$, $x_2$ $>>$ $L d(N1, Segment) = (x_{1}+x_{2})/2$ . The angle $\Phi$
is computed through geometrical relation, as follows:

\begin{eqnarray}
\Phi = (x_{2}-x_{1})/2
\end{eqnarray}
Let the coordinates of the receiver at the position $Ao$ be
$(x0, y0)$, and the coordinates of the sender $N1 (xs, ys)$.
Because of $cos(x) = cos(-x)$, two localizations of the sender
are possible, verifying the equation $\Phi = ((x2-x1)/L)$. Thus:

\vspace{0.3cm}

$x_{s} = x_{0} + d(N1, Segment)sin\Phi$\\
$y_{s} = y_{0} + d(N1, Segment)cos\Phi$

or

$x_{s} = x_{0}- d(N1, Segment)sin\Phi$\\
$y_{s} = y_{0} + d(N1, Segment)cos\Phi$

\vspace{0.3cm}
To be able to decide which position to choose for the
$N1$, the $Segment$ nodes measure the received signal
strength from the $N1$, in the direction of one of the two
localizations. Depending on the increase or the decrease of
the received signal strength, the $Segment$ nodes decides
which localization to choose for the $N1$.
Next, we determine $pdf$ of the distance of a node. Let
$p^\prime_{x(t)}(r)$ denote this $pdf$, then, given that the node has arrived
at the time $t$, then,

\begin{eqnarray}
p^\prime _{x(t)}(r) = \frac {p_{x(t)}(r)}{ P_{x(t)}(r)} , for, r < R
\end{eqnarray}

where $p_{x(t)}(r)$ and $P_{x(t)}(r)$ are shown by the eq. (9) \& eq. (10)
respectively. Since the arrival time of
a node is uniformly distributed over the interval $(0, t)$, the
unconditional density is given by, Now eq. (21) becomes as

\begin{eqnarray}
p^\prime_{x(t)}(r) =\frac{1}{t}\int_{0}^{t}\frac {p_{x(z)(t)}(r)}{P_{x(z)(t)}(r)}dz
\end{eqnarray}

Letting $p^\prime_x(t)(r)$ as the corresponding steady-state
$pdf$,then,

\begin{eqnarray}
p^\prime_{x(t)}(r)=\lim_{t\rightarrow\infty}p'_{x(t)}(r)=\lim_{n\rightarrow\infty}\frac{1}{t}\int_{0}^{t}\frac{p_{x(z)(t)}(r)}{P_{x(z)(t)}(r)}dz
\end{eqnarray}

\section{Simulations and Discussions} The simulation scenario consist of Highway model involving Vehicles moving in two directions in the same way as it happens in four-lane real Highway environment. The simulation are performed with two Mac layer protocols 802.11 and 802.11p. DSDV, DYMO and OLSR are used as routing layer protocols with both Mac layer protocols. The comparison of all these protocols is done by varying the mobility and density of Vehicles. The performance metrics used are shown in Table. 1



\textit{\textbf{Throughput}}  is the measure of data received per unit time. Its unit is bytes per second (bps). In Fig. 3.a
 maximum throughput is generated by DYMO while MOD OLSR, OLSR and DSDV.
 MOD DYMO shows lowest throughput and it is not performing as well as DYMO. DYMO is a reactive protocol it can only transmits or receives the routing packets when data arrives. 
MOD DYMO's efficiency is reduced due to the fact that number of routing packets sent per second by each node is decreased.
fig.3.a having scalability  scenario DYMO is on the top once again while the other protocols shows decrease throughput.
  MOD DYMO again shows throughput lower than that of original DYMO because DYMO is allowed to use higher bandwidth.
Nodes are mobile at a constant speed of $15m/s$ for all scalabilities scenarios.

\small
\vspace{0.05cm}

\begin{table}[!h]
\caption {Simulation Parameters}
\begin {center}
\begin{tabular}{|c|c|}
\hline
\textbf{Parameters} & \textbf{Values}\\
\hline

Simulator & NS-2(Version 2.34)\\

\hline

Channel type & Wireless  \\
\hline

Radio-propagation model  & Nakagami  \\
\hline

Network interface type & Phy/WirelessPhy, Phy/WirelessPhyExt  \\
 \hline

MAC Type  &Mac /802.11, Mac/802.11p  \\

\hline
Interface queue Type & Queue/DropTail/PriQueue \\
\hline

Bandwidth & 2Mb \\
\hline

Packet size & 512B  \\

\hline

Packet interval & 0.03s\\
\hline

 Number of mobile node & 25 nodes, 50 nodes, 75 nodes,100 nodes \\
\hline

Speed & 2 m/s,7 m/s,15 m/s,30 m/s\\
\hline

Traffic Type & UDP, CBR \\
\hline

Simulation Time & 900 s \\
\hline

Routing Protocols & DSDV, DYMO, OLSR, MOD DSDV, \\
&  MOD OLSR, MOD DYMO \\
\hline

\end{tabular}
\end{center}
\end{table}

MOD OLSR and OLSR both produce good amount of throughput because of their proactive nature and ability of decision made by each node; each node decides route up to next two hops. MOD OLSR performed slightly better because it's NRL is increased by increasing the number of control packets. Increase in the number of control packets is achieved by reducing the time interval for $HELLO$ and $TC messages$.
DSDV did not produce enough throughput.
 It happens because the topology is varying quickly and node find a 
 
\begin{figure*}[!t]
 \centering
\subfigure[MANETs Throughput vs Scalability]{\includegraphics[height=5 cm,width=7 cm]{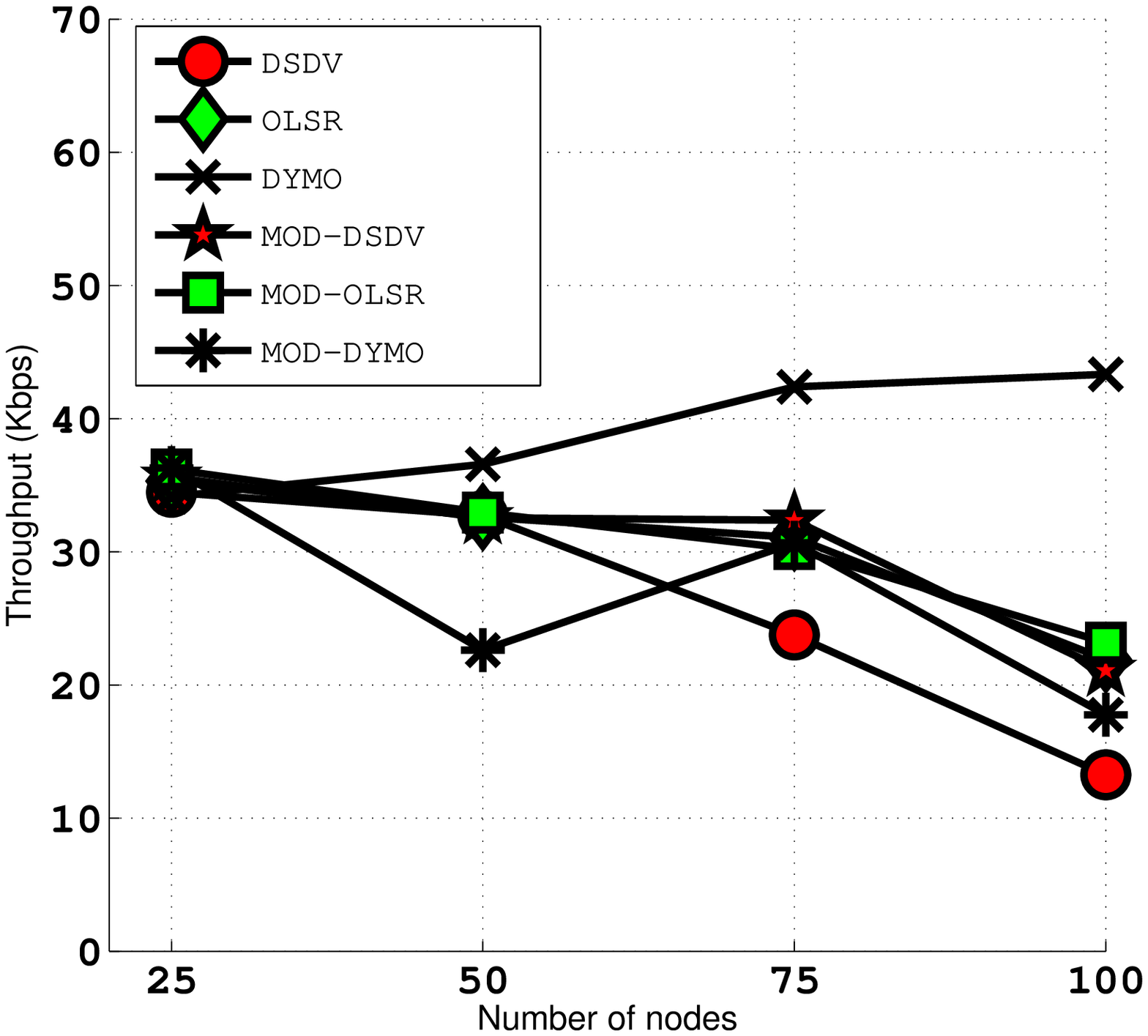}}
 \subfigure[MANETs Throughput vs Mobility]{\includegraphics[height=5  cm,width=7 cm]{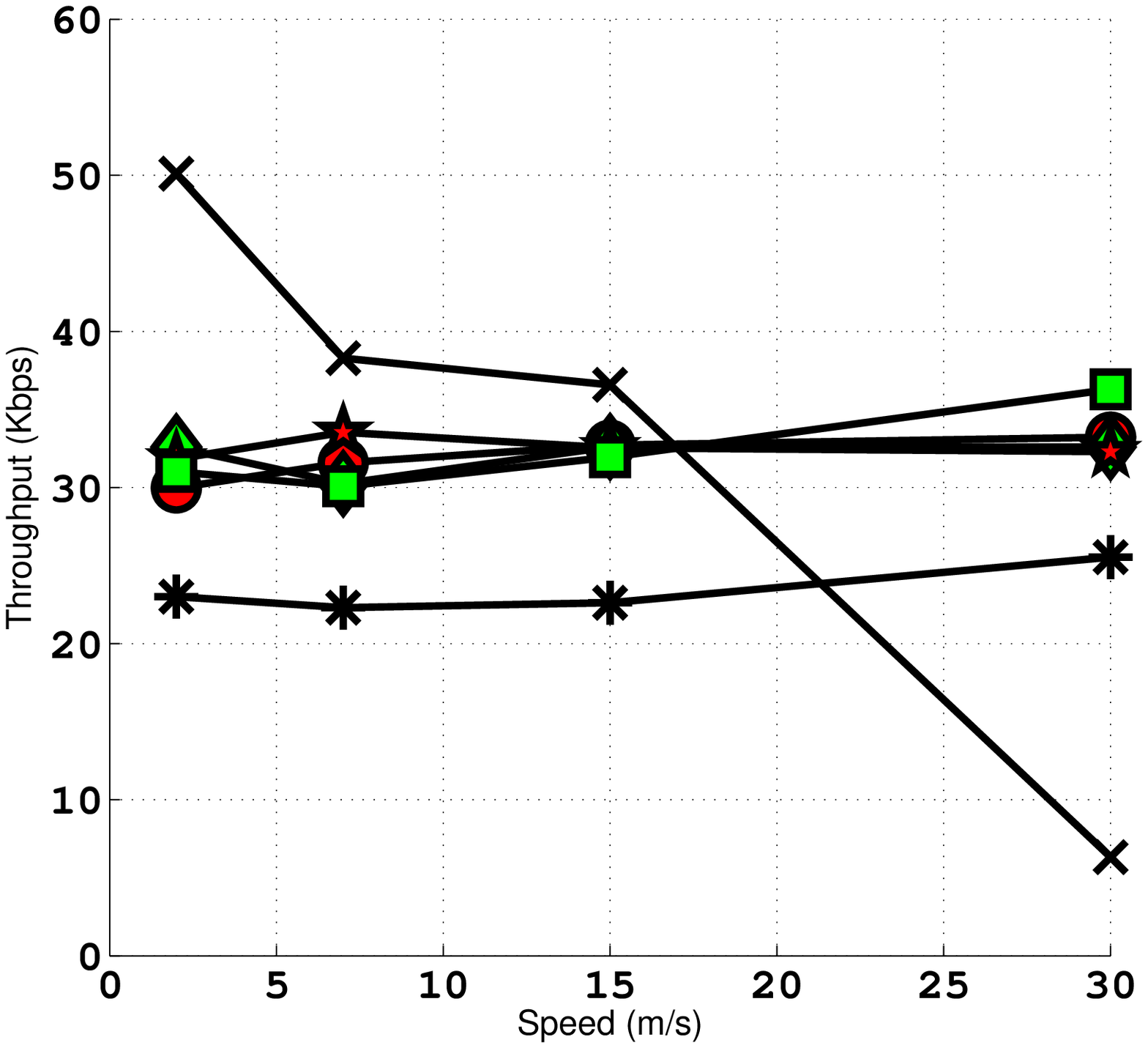}}
\subfigure[VANETs Throughput vs Scalability ]{\includegraphics[height=5 cm,width=7 cm]{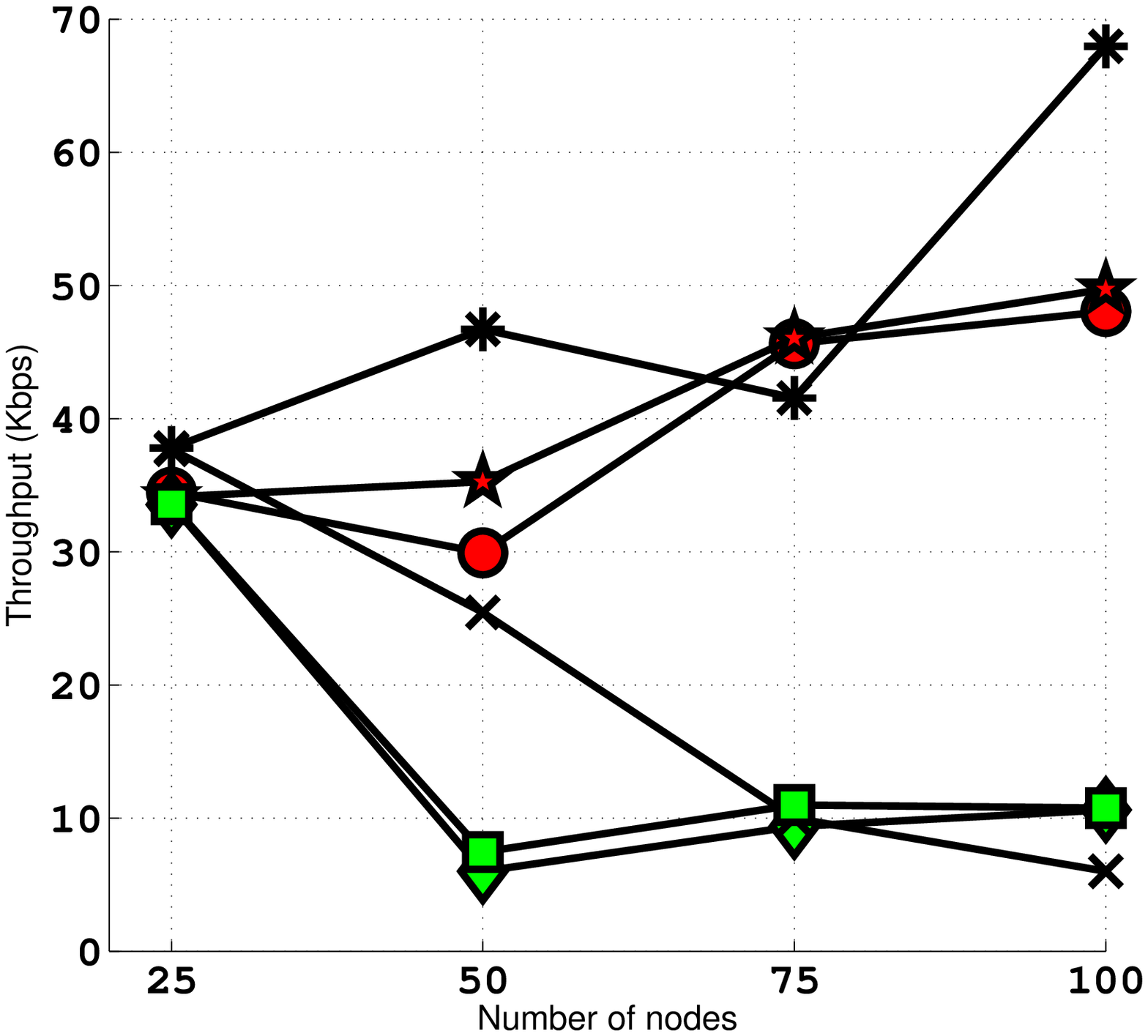}}
\subfigure[VANETs Throughput vs Mobility ]{\includegraphics[height=5  cm,width=7 cm]{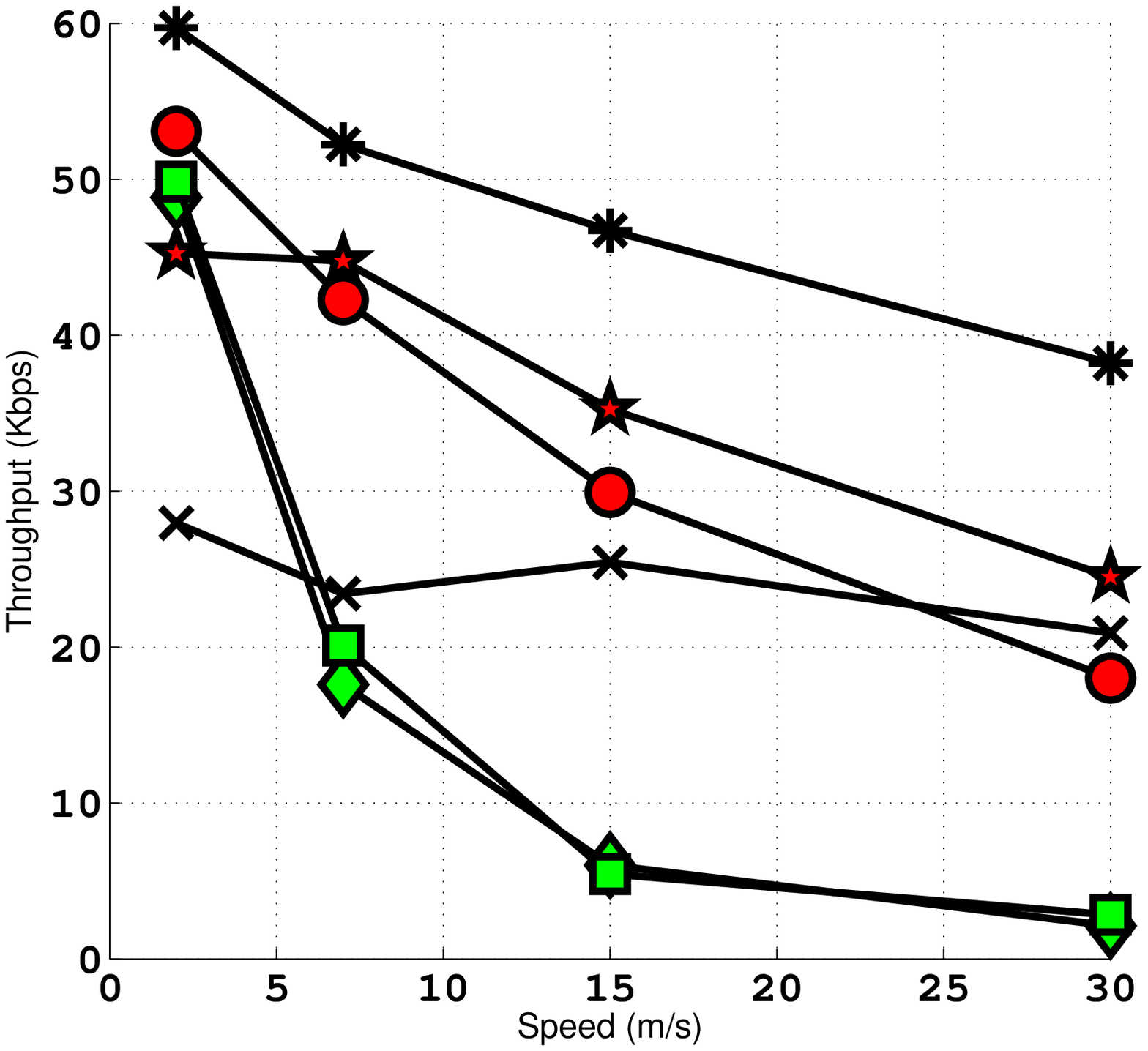}}
\caption{Throughtput}
\end{figure*}
 
route too often to send the packets. But when we increased the time between trigger updates, periodic updates and periodic update interval these issue was resolved and more packets are sent to achieve higher throughput.
In Fig. 3.b for low mobility DYMO again produce the maximum throughput while OLSR, DSDV, MOD DSDV, OLSR MOD and MOD DYMO produced in descending order. For high mobility the maximum throughput is generated by the two of proactive protocols DSDV and OLSR. MOD OLSR produces the highest amount of throughput while DSDV, OLSR and MOD DSDV also produces good amount of throughput. Due to their proactive nature and the trigger updates by DSDV and MOD DSDV and $HELLO$ plus $TC messages$ allows the DSDV and OLSR to know the state of route. Due to which the packets are more likely to reach their destination.

In Fig. 3.c in VANETs for low scalability MOD DYMO has produced the highest throughput when used with VANETs. The DYMO depends on link layer's feedback for activation and deactivation of routes. Since $802.11p$ is better than $802.11$ therefore MOD DYMO produces better throughput.
MOD OLSR slightly better than OLSR because of more proactiveness. From Fig. 3.c we can observe that
   after getting some improvements in MOD DSDV and in throughput of MOD OLSR. DYMO has been producing the minimum throughput because of high route timeout which means a useless route is stored for long period of time. MOD DSDV improved the performance of DSDV because of decreased robustness and proactiveness.

In low mobility MOD DYMO outperformes all other protocols as observed from Fig. 3.d although DYMO is been acting decently but in the given scenario the MOD DYMO proved its worth due to its smaller route timeout interval and decreased in RREQ wait time.
 Due to smaller route timeout the need of finding a new route is increased which means each route is valid (possibly). Also wait time decreases means more number of RREQs that leads to increased possibility of finding a new route.
 DSDV is working well for low mobility because of proactive nature while MOD DSDV works reasonable but not better than DSDV. OLSR also works well but MOD OLSR is working better due to more number of $HELLO$ and $TC$ $ messages$.

When the case of high mobility is taken, the MOD DYMO is again performing best. While DYMO itself is $50\%$ less efficient than MOD DYMO because of high mobility the possibility of maintaining a route for long period of time is difficult therefore DYMO is the one with less throughput than MOD DYMO.
  OLSR and MOD OLSR both are underachievers due to very proactive nature.

\section{Conclusion}
	 This paper presents path loss model along with framework for probability distribution function for VANETs. Routing protocols DSDV, DYMO and OLSR were compared for MANETs and VANETs. Our stimulation work found that overall DSDV performs fine for throughput i.e., maximum number of packets reached their destination successfully. DYMO and OLSR were underperforming and gave below throughput. DSDV and OLSR being proactive protocol stores the route as routing table entries to all destinations therefore they have the minimum E2ED while DYMO is a reactive protocol event then it worked fine when simulated with 802.11 but its performance became degraded when the Mac protocol was changed to 802.11p. When we considered the NRL, DYMO and OLSR are the ones with high NRL.
Besides the evaluating the performance of DSDV, OLSR and DYMO we also made some modifications to these routing protocols and observed their performance at the end we came up with the result that with minor changes better results can be achieved in at least one parameter.

\end{document}